**T.A. Movsessian*[1], T. Yu. Magakian*[1], A. V. Moiseev*[2]**

# Long-Slit Spectroscopy Of Herbig-Haro Outflow System, Associated With IRAS 01166+6635

1) *Byurakan Astrophysical Observatory, NAS Armenia, Aragatsotn reg., Armenia*
**2)** *Special Astrophysical Observatory, RAS, N.Arkhyz, Karachaevo-Cherkesia, Russia*

E-mail: tigmov@bao.sci.am

## 1. Introduction.

Herbig–Haro (HH) objects are compact emission nebulae characterized by shock-excited emission lines. In the majority of cases, they are formed by the interaction of collimated outflows from young stellar objects with the interstellar medium. HH objects often represent the brightest knots within collimated jets and correspond to regions, where the outflow impacts the surrounding ambient cloud material, forming so-called terminal working surfaces. In addition, shock structures may arise from velocity variability within the outflow itself, producing internal working surfaces along the jet axis. It is considered proven that the presence of HH objects is one of the main signs of an active star formation in dark clouds [1].

On the other hand, collimated outflows are, in the majority of cases, associated with the small reflection nebulae that are visible both in the optical and in the infrared. The morphology of these nebulae results from the scattering of stellar radiation that escapes preferentially along the polar directions, as defined by a circumstellar disk or toroidal envelope. The presence of such so-called cometary nebulae is therefore considered an indicator of stellar youth.

The outflow from the infrared source IRAS 01166+6635 was discovered during the Byurakan Narrow Band Imaging Survey (BNBIS), carried out with the 1-m Schmidt telescope of the Byurakan Observatory [2]. The source is located near the center of the isolated dark cloud TGU 829 [3] or Dobashi 3782 [4] with estimated distance of about 830pc (see [2] for the discussion of the distance). It is associated with a faint cone-shaped reflection nebula oriented in an N–S direction, 10″ in length. It can be seen on the PanSTARRS survey images as a very faint red star. The nebula is also visible in the near-IR range, with smaller dimensions [5].

This outflow includes the bright Herbig-Haro object, numbered as HH 1227 C, which can represent the terminal working surface of the flow [2]. In addition, two emission knots were detected: one is located at a projected distance of approximately 12″ from the source, and the second one is a compact jet-like feature in the immediate vicinity of the driving source, which is not visible in the optical range.

[1] These authors contributed equally to this work.

## 2. Observations.

To better understand the nature of this flow, we performed its long-slit spectral observations. The spectra were obtained on 21 January 2025 with the 6-m BTA telescope using the SCORPIO-2 multimode focal reducer at the prime focus [6]. The VPHG1200@540 grism was used, providing a spectral range of 3650–7200 Å at a resolution of approximately 5.5 Å. The spectrograph slit width was 1″ with length about 6.3′ and a spatial scale of 0.4″ per pixel. An EEV 261-84 CCD detector was used. The total exposure time was 3600 s, with stellar image sizes (seeing) of about 1″. The data reduction was performed in a standard way using the IDL-based pipeline for reducing long-slit spectroscopic data obtained with SCORPIO-2 [7].

The position angle of the spectrograph slit was approximately 7°; it passed along the flow through the head of the reflection nebula and HH 1227C object. The slit locations are shown in Fig. 1.

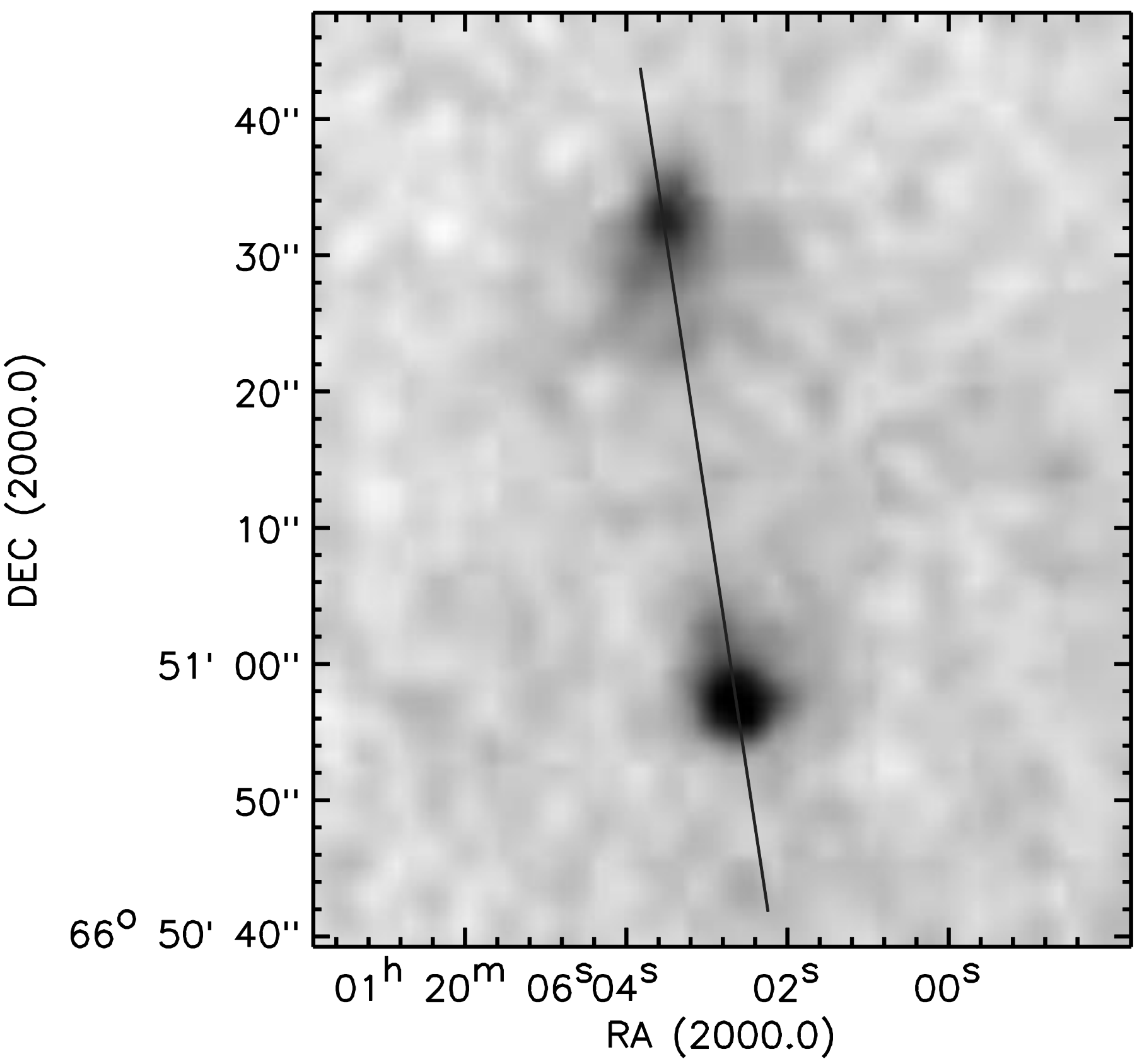

Fig.1. An image of IRAS 01166+6635 and HH 1227 flow, obtained in Hα narrow-band filter on the Byurakan 1-m Schmidt telescope.

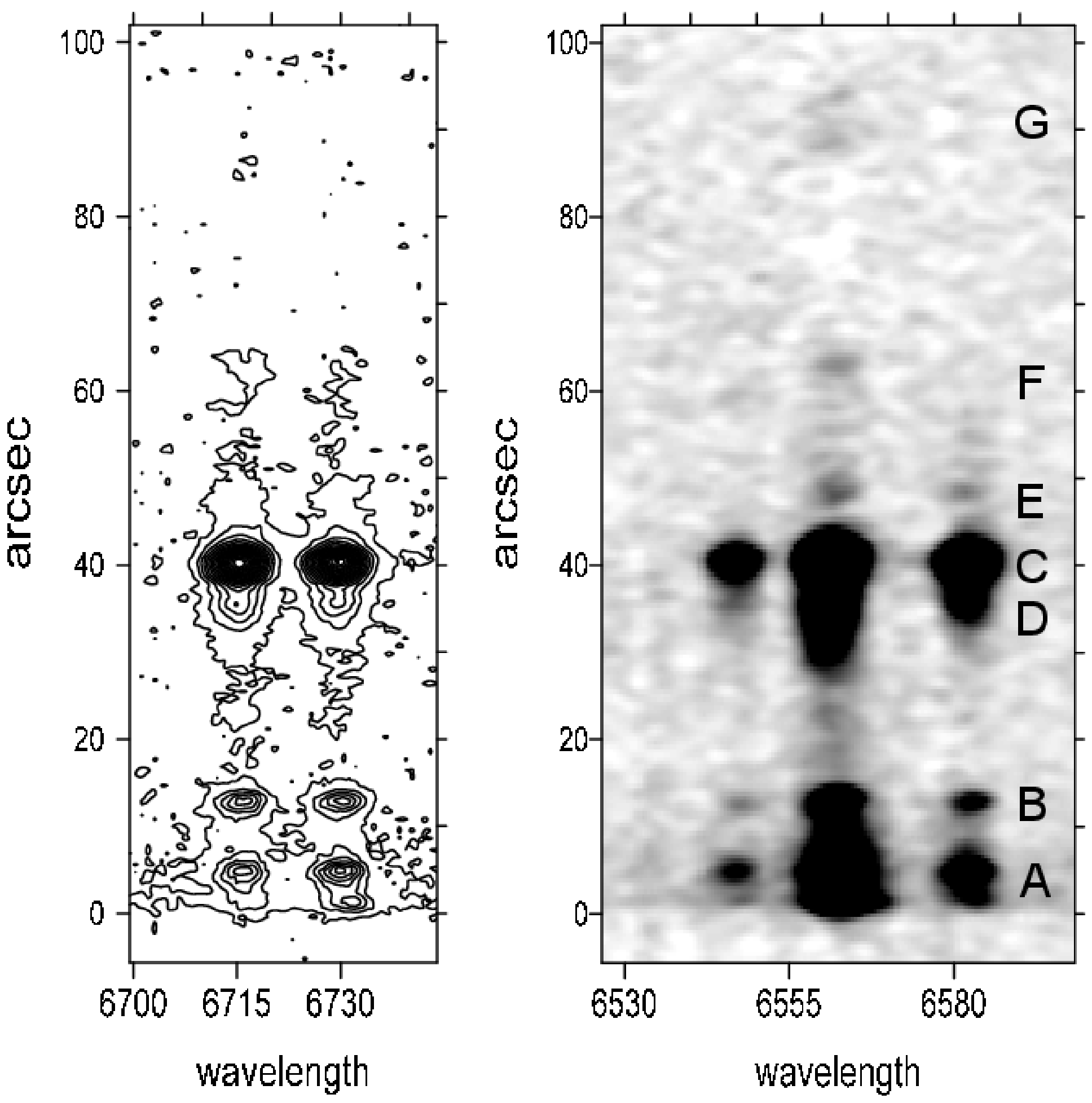

Fig.2. Knots of HH 1227 flow, visible in the long-slit spectrum of SAO 6-m telescope. Left panel – isophotes of the red doublet of [SII] emission lines. Right panel – grayscale representation of Hα and [NII] emissions. Knots are marked by letters.

### 3. Results and discussion.

In the discovery paper [2] only three HH knots were described in the HH 1227 flow. However, long-slit observations revealed many additional details. Actually, the faint emission fills all the space between the source and the brightest in the flow optical knot C and extends even beyond. Besides of knots A, B and C four more well-defined knots were found on the long-slit spectral images, making the total number of the knots in HH 1227 equal to seven. We denoted the new knots by letters from D to G. They all can be seen in the intensity grayscale representation of the spectrum in the vicinity of the Hα and [N II] lines, shown in the right panel of Fig.2. In the left panel of this figure the appearance of [S II] lines in the form of isophotes is shown. The discovery of new knots allows estimating the total extent of the system to be at least 90″, which corresponds to a projected length of ≈ 0.36 pc for the HH 1227 outflow (for the 830 pc distance). Also, on our spectrogram a very faint red stellar-like continuum with emission lines superimposed is seen near the northern edge of the continual spectrum of the reflection nebula. We interpret it as the spectrum of the source star. Besides of the usual set of the emission lines of Hα, [N II], [S II] and [O I], one can discern faint, but definitely existing emissions of [Fe II] and even λ7064 He I. We should note also, that all emission lines in the spectra of the knots and the star as well have single profiles, with FVHM (full width at half-maximum) about 5.5-6.5 Å.

As can be seen from Fig.2, the emission lines of the jet are blueshifted, but no significant velocity variations exist along the flow. Also no evidence of a counterjet is detected in this system. We analyzed the spectra of all knots and measured their radial velocities. For this we used mainly the five most intense lines: Hα, [N II] and [S II]. We excluded [O I] emissions, which are rather bright in the knots A, B and C, because they were seriously blended with night sky lines. For very faint knot G we used only Hα. The richest and most high-excitation spectrum has knot C, where, in addition to usual five intense red lines, we detected such lines as Hβ, Hγ, [O III], [N I], λ5755 [N II], λ7064 He I, λ7155 and other, more faint lines of [Fe II]. We show it spectrum on Fig.3.

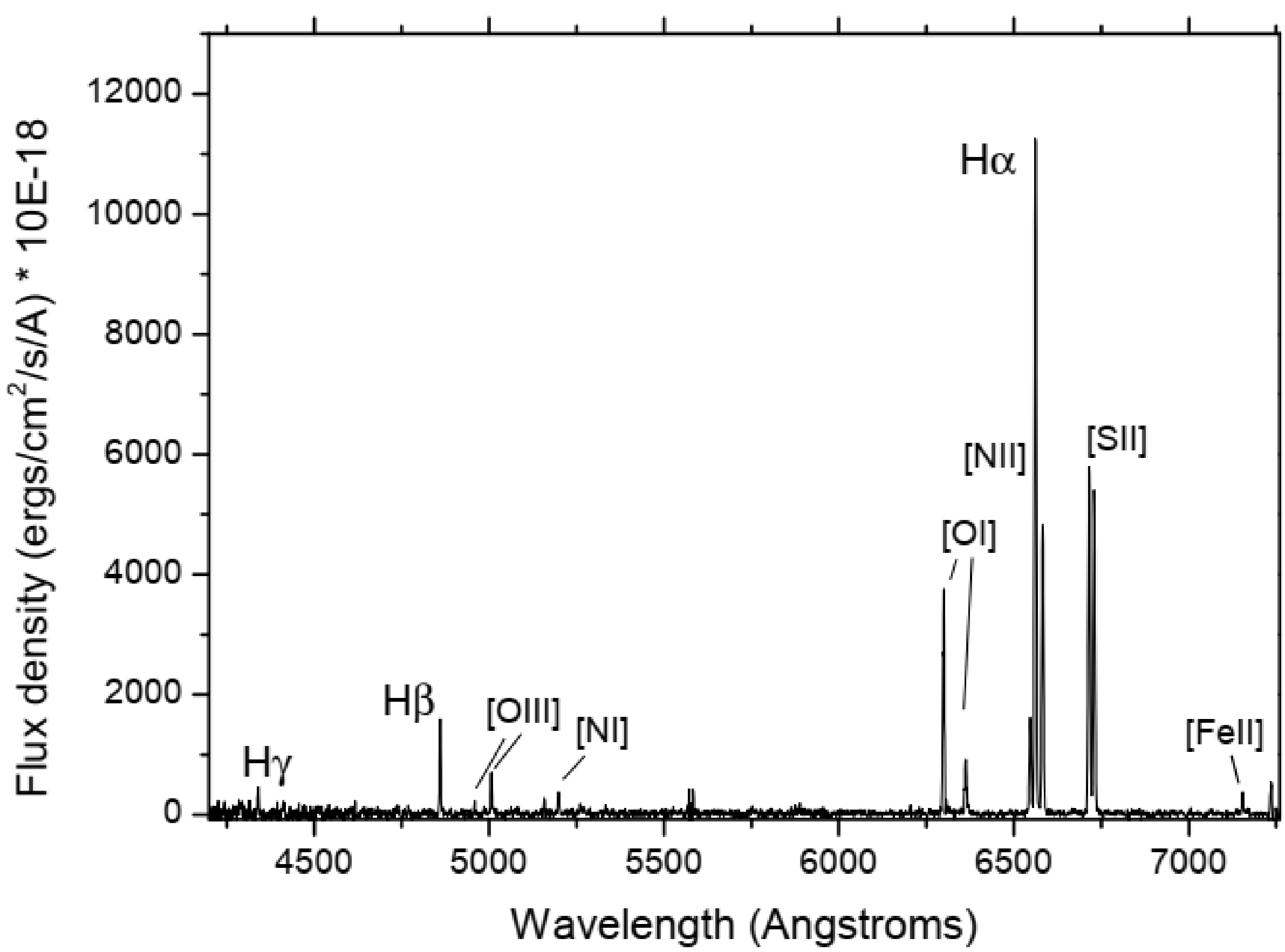

Fig.3. The spectrum of the knot C in the HH 1227 outflow. Major emission lines are labeled.

The full list of knots and their parameters is given in Table 1, where the distance of the knots from the source along the outflow, the heliocentric radial velocities and electron densities of the knots, estimated by the ratio of the [SII] doublet lines for the $T_e = 10^4$ are presented.

**Table 1.**

Physical parameters of knots in IRAS 01166+6635 outflow.

| Knot | Distance (″) | Vr (km/s) | $n_e$ ($cm^{-3}$) |
|---|---|---|---|
| source | 0 | –6 ± 8 | 8000 |
| A | 3.4 | –55 ± 7 | 3300 |
| B | 14.1 | –52 ± 1 | 450 |
| D | 34.2 | –81 ± 5 | 100 |
| C | 39.4 | –73 ± 5 | 550 |
| E | 47.8 | –70 ±14 | 180 |
| F | 61.7 | –34 ±14 | <100 |
| G | 87.3 | –70: | |

The Table 1 confirms that, as was mentioned above, the HH 1227 outflow keeps more or less same velocity along its length and probably propagates along the straight line. To estimate its inclination we can assume that the real space velocity of the outflow is near the typical value for the stellar collimated flows (about 200 km/s). Then, taking the mean radial velocity as –60 km/s, one can get the angle between the axis of the outflow and line of sight: $\approx 70°$. Thus, the real length of the HH 1227 flow will be about 0.38 pc, and its kinematic age for the distance of 830 pc can be estimated as 2000 years. As can be seen from the Table 1, the mean distance between the knots in this flow can be taken as 10″. Assuming the hypothesis, that the knots in collimated flows are produced during the outbursts of the source star (see, e.g., [8]), one can suppose that the time between the outbursts of IRAS 01166+6635 should be about 200-250 years.

These estimates should be compared with the previous observational data about this outflow [9]. In this work the radial velocity of HH 1227 C was found to be -75 km/s, in nearly perfect agreement with our measurements. But the most interesting fact is that the authors of the article [9] successfully measured proper motion of the knot C, and found that its vector is oriented directly along the axis of the outflow (p.a. = 190°), and the PM velocity was estimated as 92.4 ± 3.5 km/s. Then they found the inclination of the velocity vector as 51°. However, they used 240 pc as the distance estimation; if we apply our distance estimate 830 pc, the PM velocity will increase to 318 km/s, and the angle between the axis of the outflow and line of sight will become 77°. As one can see, these values are near to our *ad hoc* estimates, and in any case 92 km/s are too low value for the jet propagation velocity. The kinematic age estimate with new PM velocity will lower to 1250 years. Thus, we can assume that all data support the assessment of a more distant location of the object.

From the Table 1 we also see that the electronic density in the HH 1227 decreases more or less smoothly along with the distance from the star, with exception of the knot C. This knot is brightest and has also highest excitation, which points to the collision with the interstellar medium. One can suggest that the knot C can represent the terminal working surface of the flow; however, ahead of it we see several more knots. Thus, it can be the so-called internal working surface, which sometimes can be detected inside the collimated outflows [see, e.g., [9]).

## 4. Conclusion

Our observations fully confirmed the nature of HH flow from IR source IRAS 01166+6635 and revealed its complex structure. It is currently unknown whether there are other young stars in the isolated TGU 829 dark cloud, but, obviously, star formation has already started there. In our previous work [2] it was shown, that the minimal bolometric luminosity of IRAS 01166+6635 is 0.6 $L_{\odot}$. However, this estimation assumes 240 pc as the minimal distance for the source [10]. For the distance of 830 pc, which is based, though indirectly, on the measurements from Gaia DR3 catalogue, this value should be increased up to 7 $L_{\odot}$. Such luminosity is quite typical for T Tau stars. More reliable information about the nature of the source star can be obtained by conducting infrared spectroscopy.

**Acknowledgements.**

Observations at the telescopes of the Special Astrophysical Observatory of the Russian Academy of Sciences (SAO RAS) are carried out with the support of the Ministry of Science and Higher Education of the Russian Federation. The upgrade of the instrumental base is performed within the framework of the national project *Science and Universities*. The acquisition and reduction of the spectroscopic data were conducted as part of the state assignment of SAO RAS approved by the Ministry of Science and Higher Education of the Russian Federation.

**T.A. Movsessian, T. Yu. Magakian, A. V. Moiseev**

## Long-Slit Spectroscopy Of Herbig-Haro Outflow System, Associated With IRAS 01166+6635

We present long-slit spectroscopic observations of the outflow associated with an infrared source IRAS 01166+6635, conducted with the 6-m telescope of the Special Astrophysical Observatory using the SCORPIO-2 focal reducer. The structure of the flow as investigated in detail, its position-velocity diagrams are constructed. The electron densities of the several knots in the outflow were estimated, revealing a decrease in its density with increasing distance from the driving source.

**Т.А. Мовсесян, Т.Ю. Магакян, А.В. Моисеев**

## Спектроскопия с длинной щелью потока Хербига-Аро, связанного с IRAS 01166+6635

Представлены результаты длиннощелевых спектроскопических наблюдений коллимированного истечения, связанного с инфракрасным источником IRAS 01166+6635, выполненных на 6-метровом телескопе Специальной астрофизической обсерватории с использованием фокального редуктора SCORPIO-2. Подробно исследована структура потока; обнаружено несколько новых узлов. Установлено, что полная длина потока в проекции составляет около 0,36 пк, а его кинематический возраст оценивается менее чем в 2000 лет. Кроме того, были оценены электронные плотности нескольких узлов потока, что выявило уменьшение плотности по мере увеличения расстояния от источника, создающего данный поток.

Տ.Հ. Մովսիսյան, Տ.Յու. Մաղաքյան, Ա.Վ. Մոիսեեվ

## IRAS 01166+6635 ինֆրակարմիր աղբյուրի հետ կապված ուղղորդած արտահոսքի երկար ճեղքով սպեկտրոսկոպիա

Ներկայացվում են IRAS 01166+6635 ինֆրակարմիր աղբյուրի հետ կապված արտահոսքի երկար ճեղքով սպեկտրոսկոպիկ դիտումների արդյունքները, որոնք իրականացվել են Հատուկ աստղաֆիզիկական աստղադիտարանի 6 մետրանոց աստղադիտակով՝ SCORPIO-2 ֆոկալ ռեդուկտորի կիրառմամբ: Մանրամասն ուսումնասիրվել է հոսքի կառուցվածքը, և հայտնաբերվել են մի քանի նոր հանգույցներ: Պարզվել է, որ հոսքի ընդհանուր պրոյեկցված երկարությունը կազմում է մոտ 0.36 պարսեկ, իսկ նրա կինեմատիկական տարիքը գնահատվում է 2000 տարուց պակաս: Բացի այդ, գնահատվել են արտահոսքի մի շարք հանգույցների էլեկտրոնային խտությունները, ինչի արդյունքում բացահայտվել է, որ խտությունը նվազում է աղբյուրից հեռավորության մեծացման հետ:

---

Author Contributions: authors contributed equally to this work.

Data Availability Statement: The authors confirm that the data supporting the findings of this study are available within the cited articles.

Conflicts of Interest: The authors declare no conflict of interest.